\documentclass{article}
\usepackage{arxiv}
\usepackage[utf8]{inputenc}
\usepackage[T1]{fontenc}
\usepackage{hyperref}
\usepackage{booktabs}
\usepackage{graphicx}
\usepackage{natbib}
\usepackage{multirow}

\title{Signal vs Noise in Eye-tracking Data: Biometric Implications and Identity Information Across Frequencies}

\author{ {\hspace{1mm}Mehedi Hasan Raju}\thanks{corresponding author} \\
	Texas State University\\
    601 University Drive\\
    San Marcos, Texas, 78640, USA\\
	\texttt{m.raju@txstate.edu} \\
	\And
	{\hspace{1mm}Lee Friedman} \\
	Texas State University\\
    601 University Drive\\
    San Marcos, Texas, 78640, USA\\
	\texttt{lfriedman10@gmail.com} \\
	\And
	{\hspace{1mm}Dillon J. Lohr} \\
    Texas State University\\
    601 University Drive\\
    San Marcos, Texas, 78640, USA\\
	\texttt{djl70@txstate.edu} \\
	\And
	{\hspace{1mm}Oleg V. Komogortsev} \\
    Texas State University\\
    601 University Drive\\
    San Marcos, Texas, 78640, USA\\
	\texttt{ok@txstate.edu} \\
}

\date{}

\renewcommand{\shorttitle}{Signal vs Noise in Eye-tracking Data}

\begin{document}
\maketitle

\begin{abstract}
Prior research states that frequencies below 75~Hz in eye-tracking data represent the primary eye movement termed ``signal''
while those above 75~Hz are deemed ``noise''. 
This study examines the biometric significance of this signal-noise distinction and its privacy implications. 
There are important individual differences in a person's eye movement, which lead to reliable biometric performance in the ``signal'' part.
Despite minimal eye-movement information in the ``noise'' recordings, there might be significant individual differences.
Our results confirm the ``signal'' predominantly contains identity-specific information, 
yet the ``noise'' also possesses unexpected identity-specific data.
This consistency holds for both short-($\approx$ 20 min) and long-term ($\approx$ 1 year) biometric evaluations. 
Understanding the location of identity data within the eye movement spectrum is essential for privacy preservation.

\end{abstract}

\keywords{eye movement, biometric, signal, noise}

\maketitle

\section{Introduction}
\begin{figure*}[htbp]
\centering
\includegraphics[width=0.9\textwidth]{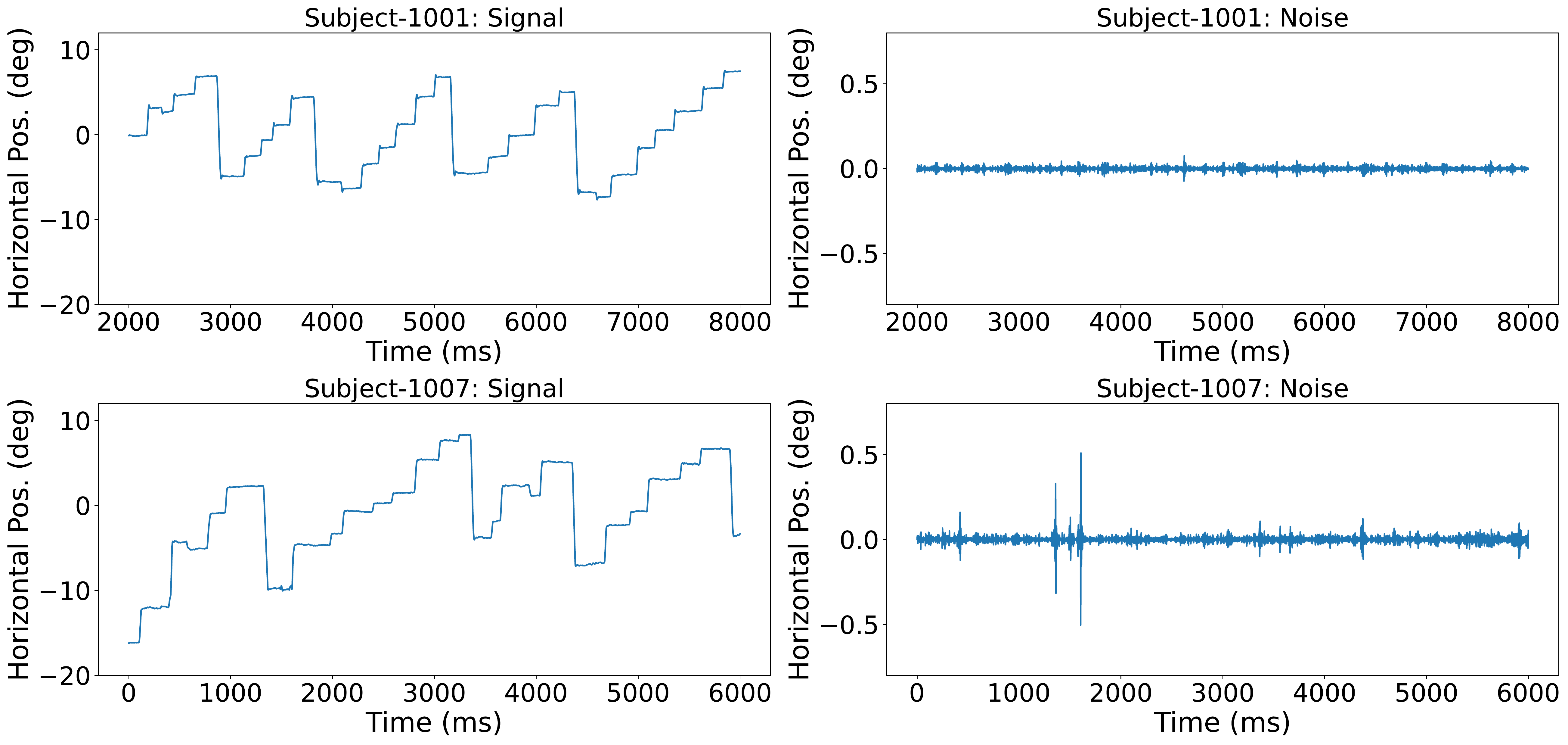}
\caption{Signal and Noise comparison: Exemplars. The two plots on the left column of the figure are the raw data collected at 1000 Hz. We refer to these recordings as the ``raw'' data.  The two plots in the middle column of the figure are recordings after a low-pass filter was applied. We refer to these recordings as the ``signal'' portion of the recording.  The two plots on the right column are high-pass filtered recordings from the same subjects.  We refer to these as the ``noise'' portion of the recording. Each row is a different subject.}
\label{fig:comparison}
\end{figure*}

A person's identity is reflected in both their biological and behavioral characteristics. 
In ordinary life, we recognize our friends and colleagues by such characteristics.  
It is well established that anatomical characteristics such as face, fingerprint, and iris can be used to authenticate individuals in biometric systems\citep{jain2007handbook,biometric2010}.  
Similarly, behavioral characteristics such as handwriting, voice, and signature can be used in the same way.  
Currently, there is substantial interest in eye movement based biometric analysis.  
The human oculomotor system is a highly complex amalgam of neurological, physiological and anatomical systems.  
Recent biometric authentication studies have been based on oculomotor recordings \citep{lohr2020metric,lohr2022eye, Lohr2020, lohrTBIOM, deepeyedentification, deepeyedentificationlive}.  
Eye movements have also been shown to be spoof-resistant \citep{rigas2015,raju2022iris, Komogortsev2015}. 
Several research teams are currently working on eye movement-based biometrics (EMB) to create an advanced machine-learning approach that can be applied in practical situations.
Typically,  EMB studies employ either filtered or unfiltered signal types.  
We are not aware of any prior EMB research that has tried to split eye-movement data into its ``signal'' and ``noise'' portions using digital filters.
See Fig. \ref{fig:comparison} for several exemplars of ``signal'' and ``noise''. 

In addition to biometric performance, our study has implications for the privacy of an individual's identity. 
Studies have investigated that, the eye movement of a person contains individual-specific information such as race, ethnicity \citep{ethnicity1,ethnicity2}, age \citep{age, cantoni2015gant}, and gender \citep{gender, sammaknejad2017gender}. 
From a privacy perspective, these are highly sensitive individual information \citep{rui2018survey}. 
Suppose we can understand which part (sine-wave frequencies) of the individual's eye movement data is responsible for biometric performance enhancement. In that case, we might be able to understand which part is required to be hidden for privacy concerns.
If the goal is to reduce identity-specific information in the form of encoded identity, it is important to determine whether it is primarily encoded in the eye movement ``signal'' itself or in the person-specific ``noise''. 

This study does not aim to reduce person-specific ``noise''; instead, its main research question is whether identify-specific information resides in the ``signal'' part or the ``noise'' part. The primary objectives of this research are:
\begin{itemize}
\item    To distinguish between the ``signal'' and ``noise'' portions of eye movement data separately using established digital filters \citep{raju2023filtering}.
\item    To compare the biometric performance on the ``signal'' and ``noise'' parts of the eye movement data. 
We hypothesize that the ``signal'' part of the recordings will perform better than the ``noise'' part but we are uncertain how the ``noise'' part will perform. 
If the ``noise'' part carries identity-specific information, it might perform significantly better than only device-specific noise.
\item    To compare biometric performance on ``signal'' and ``noise'' from a short-term dataset (recordings $\approx$ 20 min apart) to a long-term (recordings $\approx$ 1 year apart) dataset to understand the longitudinal pattern of this phenomenon.
\end{itemize}

\section{Prior Work}

\subsection{Prior Work on EMB}
Kasprowski and Ober \citep{Kasprowski2004} introduced eye movement as a biometric modality that can be used in human identification. 
Since then, there has been much research related to eye movement biometrics. The main objective was to come up with a state-of-the-art approach for user authentication. 
The approaches can be categorized into two types: statistical feature-based approach and end-to-end machine learning-based approach.

The statistical feature-based approach involves a standard processing pipeline, where recordings are divided into discrete eye movement events using a classification algorithm, and then a biometric template is formed as a vector of discrete features from each event~\citep{Rigas2017}. 
However, event classification can be challenging~\citep{Andersson2017} and impact biometric performance, depending on the event classification algorithm used. 
Several eye-movement event classification algorithms have been proposed \citep{ONH, classification1, classification2, Friedman2018}.  
Prior research such as \citep{Lohr2020, friedman2017method, Li2018} employed this type of approach.

On the other hand, recently end-to-end deep learning workflows have become very popular.  
There are many studies of this type already published (\citep{lohr2020metric, lohr2022eye, lohrTBIOM, deepeyedentification, deepeyedentificationlive, Jia2018, Abdelwahab2019, biometricidentification}).

\subsection{Prior Work on Filtering Eye Movement Data}
Stampe \citep{stampe} proposed two heuristic filters for video-oculography data. 
These filters have been widely adopted and modified by various manufacturers, including SR-Research and SMI. SR Research (EyeLink 1000 manufacturer) uses a modified version of the original heuristic filters proposed by Stampe. 
Within the EyeLink family of eye trackers, these two filters are known as the standard (STD) and extra (EXTRA) filters.

Various digital filter types \citep{raju2023filtering, mack} have been applied to eye-movement recordings.  
In most cases, low-pass filters were applied to eliminate the ``noise'' portion of the recording.  According to \citep{raju2023determining}, lower-frequency components ($\leq 75$ Hz) comprise the ``signal'' portion of the recordings, and higher frequency components ($>75$Hz) comprise the ``noise'' portion of the data.  
These prior works motivated us to study the biometric performance of ``signal'' compared to ``noise''. 
As far as we know, there is no research conducted on classifying the noise such as device-specific noise or person-specific noise.

\section{Methodology}
\label{method}

\begin{figure*}[htbp]
\centering
\includegraphics[width=\textwidth]{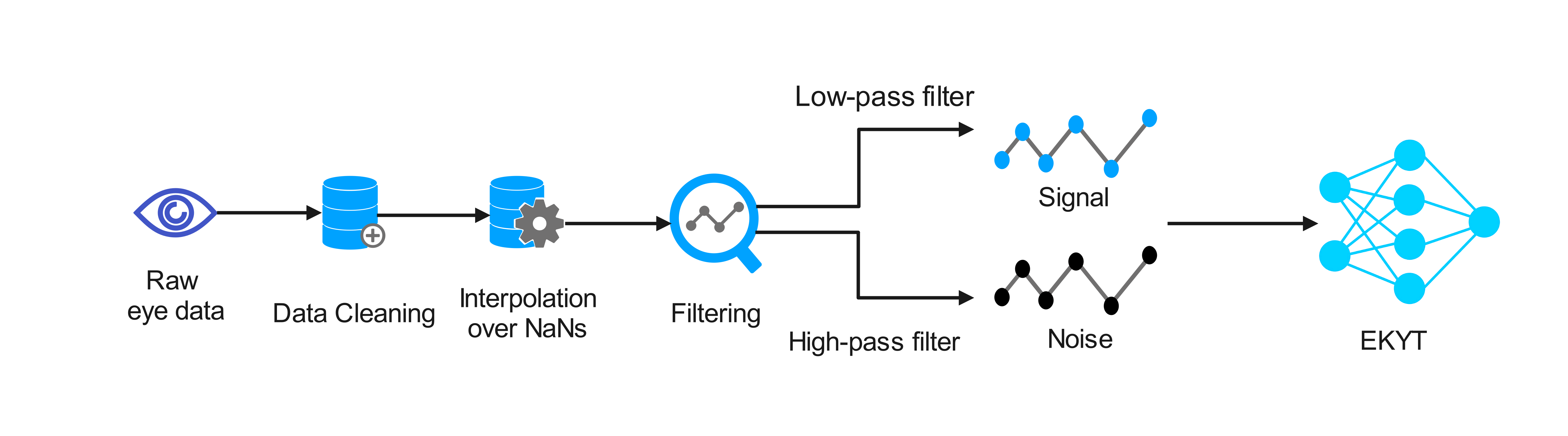}
\caption{Block diagram of the methodology}
\label{fig:method}
\end{figure*}

A basic block diagram of the methodology is presented in Fig.\ref{fig:method}.
We will be using Eye Know You Too (EKYT) network architecture \citep{lohr2022eye}, training, and evaluation methodology. 
As far as we are aware, it is the state-of-the-art end-to-end machine learning-driven architecture.

\subsection{Dataset}
The eye-movement recordings we used in this study are from the GazeBase~\citep{Griffith2020} dataset, which is publicly available.
The GazeBase dataset contains very high signal-quality eye-movement recordings as it is recorded with an EyeLink~1000 eye tracker at 1000~Hz sampling rate.
GazeBase dataset consists of 12,334~monocular~(left eye only) eye-movement recordings captured from 322~college-aged subjects. 
There are 9 rounds (Round 1 - Round 9) of recordings collected over three years.
Each recording contains the horizontal and vertical components of the left eye's gaze position in degrees of the visual angle.
Every subject went through a series of~7 eye movement tasks: a horizontal saccades task~(HSS), a fixation task~(FXS), a random saccades task~(RAN), 
a reading task~(TEX), two video viewing tasks (VD1 and VD2) and one video-gaming task (Balura game, BLG).  
Each round consists of two recording sessions of similar tasks per subject at approximately 20-minute intervals.
We present biometric analysis from two different time frames.  
The first analysis is based solely on Round 1 and compares the first session of Round 1 to the second session of Round 1.  
The second analysis compares the first session of Round 1 to the second session of Round 6.  
Although many more subjects were available for Round 1, we wanted both analyses to be based on the same number of subjects and the second analysis contained 59 subjects.  
More details about the dataset can be found in~\citep{Griffith2020}.
Throughout the paper, the term ``short-term'' refers to the data collected in Round-1 with approximately 20-minute intervals between sessions. 
The term ``long-term'' refers to the data collected in Round-6, which was obtained with approximately a year interval from Round-1.

It is important to mention that the GazeBase dataset is already filtered to the ``EXTRA'' level. 
Although this filter does suppress high-frequency noise, it underperforms digital filters \citep{raju2023filtering}. 
High-frequency data are still present in the signal after the application of this filter.

\subsection{Data Preprocessing}
Prior to being fed into the network architecture, all recordings from the dataset underwent a series of pre-processing steps.  

\subsubsection{Filtering}
EyeLink 1000 cannot estimate gaze, during a blink, the device returns a Not a Number (NaN) code for the relevant samples.  
The range for the possible horizontal component of the gaze positional value is from -23.3 to +23.3 
whereas the vertical component is from -18.5 to 11.7 degrees of visual angle (dva).  
We excluded (set to NaN) any individual gaze samples where the subjects were viewing beyond the screen dimensions.  
Our filter method required that NaNs be removed, so missing data portions were replaced via linear interpolation.
Our analysis required two types of filtering: low-pass to obtain ``signal'' and high-pass to separate ``noise''.  
Both low-pass and high-pass filtering were performed with an FIR-type filter each with 79 taps.  
The cutoff frequency (-3db point, where the signal is reduced by 49.9\%) was 75 Hz for both filter types \citep{raju2023filtering}. 
After filtering, all the NaNs were reinserted into the signal recordings. 

\subsubsection{Data Preprocessing for Embedding Analysis}
Next, we created two velocity channels (horizontal and vertical) from the raw signal using Savitzky-Golay \citep{savitzkyGolayM} with window size=7 and order=2 \citep{friedman2017method}. 
Then using a rolling window method, we split recordings into non-overlapping windows of 5 seconds (5000 samples). 
It is to be noted that 12 of these 5-second sections are combined into a single 60-second data for further evaluation.

\subsubsection{Velocity Clamping}
Velocities are clamped between ±1000◦/s as a means of reducing the impact of noise on the data. 
Afterward, all the velocity channels across all segments and subjects were z-score transformed. 
Finally, replace all NaN with 0. It is part of the data handling process established by \citep{lohrTBIOM}. 
More data preprocessing details can be found in \citep{lohr2022eye}  
 
\subsection{Eye Know You Too Network architecture}
We used the state-of-the-art network architecture for eye-movement-based biometric authentication named Eye Know You Too (EKYT) \citep{lohr2022eye}.
It is a denseNet-based \citep{densenet} architecture that achieves state-of-the-art EMB performance in the authentication situation on high-quality data (data collected at 1000~Hz). 
When enrolling and authenticating while performing a reading activity, it achieves 0.58\% EER with 60 seconds of eye movements.
The network architecture has 8 convolutional layers. 
Consequently, before being input into the following convolution layer, the feature maps created by each convolution layer are concatenated with all preceding feature maps. 
A 128-dimensional embedding of the input sequence is then created using the final set of concatenated feature maps after they have been flattened and sent through a global average pooling layer and fed into a fully-connected layer.
For more details about the network architecture, we refer the reader to \citep{lohr2022eye}.

\subsection{Dataset split and Training procedure}
The model was trained on all data from Rounds 1-5 except for BLG task. 
There were 322 subjects in Round 1 and 59 in Round 6.  The 59 subjects in Round 6 are a subset of all subjects in Round 1.  
The data from these 59 subjects from both rounds were treated as a held-out dataset and not used for training and validation.
Next, we divided the remaining participants into four non-overlapping folds for cross-validation. 
The training set is divided into 4 folds, each containing distinct classes, with an aim to distribute the number of participants and recordings as evenly as possible. 
The employed fold assignment algorithm is discussed in \citep{lohrTBIOM}.
We trained 4~different models, each one using a different held-out fold as the validation set and the remaining 3~folds as the training set for both sets of filtered data (``signal'' and ``noise'' parts).
We employed the Adam\citep{adam} optimizer, and PyTorch's OneCycleLR with cosine annealing \citep{Lr} for learning rate scheduling in our training.
We used multi-similarity loss (MS) loss~\citep{Wang2019}. PyTorch Metric Learning~(PML) library~\citep{Musgrave2020a} is used to implement it.
Hyperparameters for MS loss and all other optimizer hyperparameters were left at their default values following \citep{lohr2022eye}.

Each input sample is comprised of two channels: horizontal and vertical velocity and contains a window of 5000-time steps.
The model was trained for 100 epochs. During the first 30 epochs, the learning rate is set to $10^{-4}$ and gradually increases up to $10^{-2}$. Subsequently, over the next 70 epochs, the learning rate gradually decreases down to a minimum of $10^{-7}$.
Each batch consists of 64 samples (classes per batch = 8 $\times$ samples per class per batch = 8).

\subsection{Evaluation}
Biometric performance was assessed at two-time intervals, short term (Round 1, session 1 vs session 
2, $\approx$ 20 min) and long-term (Round 1 session 1 versus Round 6 session 2, $\approx$ 1 year).
For both time intervals, each raw recording was filtered to produce a ``signal'' portion and a ``noise'' portion.   
The final biometric assessment was based solely on the reading task (TEX) recordings. 
To maintain consistency with previous EMB research, we opted for the reading task.
To generate embeddings for each window in both the enrollment and authentication sets, 
we use the four models that were trained using a 4-fold cross-validation approach. 
We computed 128-dimensional embedding for each model and concatenated them to create a single 512-dimensional embedding for each window. 
This effectively treats the four models as a single ensemble model.

\subsection{Metrics}
We used three metrics to measure and compare our performance of the model.
These are equal error rate (EER) \citep{rigas2015}, decidability index d' (d-prime) \citep{daugman2000biometric}, and false rejection rate (FRR) \citep{conrad2015cissp, Quality2012} at a fixed false acceptance rate (FAR). 

The EER is the location on a receiver operating characteristic (ROC) curve where the FRR and FAR are equal.
The lower the EER value, the better the performance of the biometric system is. 
It is an indicator of biometric performance, usually when performed for authentication/verification tasks.
To calculate the EER, we need two disjoint subsets of data: one for enrollment and another for authentication (or verification). 
In our approach, we formed the enrollment dataset by using the first 60 seconds of the session-1 TEX task from Round 1 for each subject in the test set. 
For the authentication dataset, we used the first 60 seconds of the session-2 TEX task from Round 1 for each subject in the test set.  
For the longitudinal study, we used the first 60 seconds of the session-2 TEX task from Round 6 as the authentication dataset.
It is to be noted that we did not use 60 seconds at once, we split 60 seconds into 5-second subsequences, getting embeddings for each subsequence, and then computed the centroid of those embeddings.

We computed the decidability index d' (d-prime), which is a measure of the separation of genuine/impostor similarity distributions. 
It depends on the means of the distributions and their standard deviations. 
In general, a larger d' should be connected with a lower EER because it shows a greater separation between the genuine and imposter distributions i.e. the higher the d' value, the better the biometric system is.

Complementary to these, we also used FRR. 
FRR is one of the three metrics used in biometric accuracy. 
When the biometric system rejects an authorized subject as an unauthorized one, it is called false rejection. 
FRR indicates how often legitimate users are wrongly rejected whereas FAR presents the security risk posed by imposters gaining access.
EMB needs to be competitive with existing methods. Here, the security level of a 4-digit PIN is used as a benchmark following \citep{lohr2022eye, lohrTBIOM} because it's a common security measure. Assuming all 10,000 combinations are equally likely, a 4-digit PIN offers a very low FAR of 1 in 10,000. To compare EMB to this level of security, the paper estimates the FRR of EMB when the FAR is set to a strict level of 1 in 10,000 (written as FRR @ FAR $10^{-4}$).
According to FIDO biometric requirements \citep{FIDO2020}, FRR is used to measure the biometric performance and FRR should be no more than 5\% @ FAR $10^{-4}$.

In short, EER offers a single value for the overall error rate, making it easy to compare the biometric performance of systems. 
d' will provide us with a more nuanced understanding of the separation between genuine users and imposters.
A low FRR @ FAR $10^{-4}$ would indicate the feasibility of the EMB system for real-world use.

\subsection{Hardware \& software}
All models were trained on Dell Precision 3660.
The workstation was equipped with NVIDIA GeForce RTX 3080, an Intel i5-12600K CPU @ 3.7~GHz (10~cores), and 32~GB RAM.
The compatible Anaconda environment was set up with Python~3.7.11, PyTorch~1.10.0 (Torchvision~0.11.0,  Torchaudio~0.10.0), Cudatoolkit~11.3.1, and Pytorch Metric Learning(PML)~\citep{Musgrave2020a} version~0.9.99.

\section{Result}

Table~\ref{tab:GB4Folds} presents our results. 
It provides biometric performance of ``signal'' and ``noise'' portions of the eye movement time-series for short-term ($\approx 20$ min) and long-term ($\approx$ 1 year) evaluations.
The performance metrics used were EER, d', and FRR @ FAR $10^{-4}$.  
It is clear that the biometric performance of the ``signal'' part of the eye-movement time-series is better than the unfiltered data and outperforms the performance of the ``noise'' part of the eye-movement time-series for both time intervals. 
However, it should be noted that the noise portions performed much better than chance (for EER and FRR chance is 50\%). 

\begin{table*}[htbp]
  \centering
  \caption{Comparison of biometric performance for ``Signal" and ``Noise" in both short-term and long-term evaluations. Arrows indicate whether a larger or smaller value is better. ``Signal'' means low-pass filtered data, ``Noise'' means high-pass filtered data, and ``Unfiltered'' means data that went through our data-processing steps except for the filtering steps.}
  \begin{tabular}{|l|ccc|ccc|}
  \hline
  \multicolumn{1}{|c|}{\multirow{2}{*}{Metrics}} & \multicolumn{3}{c|}{Short term}& \multicolumn{3}{c|}{Long term} 
  \\ \cline{2-7}
  
  \multicolumn{1}{|c|}{} & 
  \multicolumn{1}{c|}{Signal} & \multicolumn{1}{c|}{Noise} & Unfiltered 
  &\multicolumn{1}{c|}{Signal} & \multicolumn{1}{c|}{Noise} & Unfiltered 
  \\ \hline
  
  EER (\%) $\downarrow$ &
  \multicolumn{1}{c|}{0.56} & \multicolumn{1}{c|}{7.66} &  {0.76} 
  & \multicolumn{1}{c|}{3.68} & \multicolumn{1}{c|}{23.73} & 5.1
  \\ \hline
  
  d-prime $\uparrow$ & 
  \multicolumn{1}{c|}{4.83} & \multicolumn{1}{c|}{2.77} &   {4.82}
  & \multicolumn{1}{c|}{3.77} & \multicolumn{1}{c|}{1.51} & 3.68
  \\ \hline
  
  FRR @ FAR $10^{-4}$(STD) (\%)$\downarrow$  & \multicolumn{1}{c|}
  {6.78 (0.18)} & \multicolumn{1}{c|}{47.73 (4.7)} & {5.1 (0.16)}
  & \multicolumn{1}{c|}{31.24 (3.26)} & \multicolumn{1}{c|}{86.34 (0.47)} & 26.54 (2.16)
  \\ \hline
  
\end{tabular}
\label{tab:GB4Folds}
\end{table*}

\section{Discussion}

Our analysis separates eye movement data into ``signal'' and ``noise'' components using established digital filters \citep{raju2023filtering}. Interestingly, identity-specific information resides not only in the expected ``signal'' portion but also to a significant extent within the ``noise'' portion of the data. This finding extends to long-term data, as validated by our longitudinal study.

\subsection{Improvement of Biometric Performance upon Signal and Noise Separation}
The main finding of the present report is that the ``signal'' portion of the eye-movement time-series dramatically outperforms the ``noise'' portion and also performs better than the unfiltered data in EER and d-prime metrics. 
Unfiltered data performs better than both ``signal'' and ``noise'' in the FRR metric.
The lower EER for the ``signal'' part indicates that it is more reliable for biometric authentication. 
Similarly, the higher d' for the ``signal'' part indicates that it has a better distinctive ability between genuine and impostor subjects. 
Finally, the lower FRR for the ``signal'' part indicates that it has better accuracy in verifying genuine subjects. 
The ``signal'' portion contains eye-movement information.  
As mentioned above, eye movements during reading include fixation, saccades, and post-saccadic oscillations. 
The biometric algorithm was apparently able to extract substantial individual-specific information from this portion.

\subsection{Biometric Information in ``Noise'' Portion}
Although, the ``signal'' portion of the eye-movement outperforms the ``noise'' portion, the ``noise'' portion performs substantially better than chance for EER (For EER 50\% is the chance). 
It is unclear at this time what qualities of the ``noise'' portion are most important for biometric performance.  
Holmqvist et al \citep{holmqvist2017} have reported on a large number of factors that affect either the spatial accuracy or spatial precision of eye-movement recordings. 
For example, according to these authors, 
subjects with larger pupil sizes have lower signal quality. 
Subjects who wear glasses have lower eye-tracking quality. 
Subjects with blue eyes have lower signal quality. 
Young-aged subjects have better spatial precision in the eye-movement recordings.
Subjects with make-up (eyeliner, eye shadow, mascara) have lower spatial accuracy.
Numerous other factors are mentioned in that article. 

In future studies, it might be of interest to see if our filtering steps might improve recording quality in the presence of these and similar factors.
These factors might contribute to individual differences in the ``noise'' portion.
What is needed is a taxonomy of noise.  We know the different types of eye movements in the ``signal'' portion, but the types of noise have not been classified.
This could be a future research project.

We also studied the longitudinal effect by comparing the performance of the model on short-term and long-term data.
It is not surprising that when biometric performance is assessed over a very short time interval ($\approx$ 20 min), performance is going to be much better than over a much longer time interval($\approx$ 1 year).
It is important to remember that, with an interval of more than a year, eye movement-based biometric authentication is impressive \citep{lohr2022eye}.

\subsection{Implication for Sampling Frequency}
We conclude that eye-movement signal contained in frequencies from 0 to 75 Hz has implications for the proper sampling rate for future studies.
Most researchers use the Nyquist theorem and conclude that for the analysis of \textit{F Hz} signals, the sampling rate needs to be \textit{2$\times$F Hz} \citep{shanon}.
But this is true only for the frequency domain.  Eye-movement time series are observed in the time domain. 
There is a common rule of thumb that, to observe a frequency of F Hz in time domain, the sampling rate needs to be \textit{10$\times$F Hz}. 
Please follow the links in the footnote for a defense of this position\footnote{1. \url{https://community.sw.siemens.com/s/article/digital-signal-processing-sampling-rates-bandwidth-spectral-lines-and-more}, 
2.\url{https://en.wikibooks.org/wiki/Analog_and_Digital_Conversion/Nyquist_Sampling_Rate} and 3. \url{https://www.dataq.com/data-acquisition/general-education-tutorials/what-you-really-need-to-know-about-sample-rate.html}}.  
Since we consider sine-waves of 75 Hz as part of the ``signal'', the minimum sampling rate required for full eye-movement preservation is 750 Hz.
Our results have implications for future studies of eye-movement biometrics.  Obviously, the signal portion provides better performance.  
In our case, with data collected at 1000 Hz, the signal portion of the data can be isolated using a low-pass filter\citep{raju2023filtering}.  
It is unclear how effectively existing methods for separating ``signal'' and ``noise'' will translate to eye-tracking data collected at much lower sampling rates.

\subsection{Implication for Privacy}
One of the critical implications of our study is the privacy concerns of the individual’s identity-specific information contained in eye-movement data. 
At this point, we can say the ``signal'' part carries the most substantial amount of information about the identity of a person. 
The ``noise'' part is also somewhat pre-filtered also carries nonrandom person-specific information.
It is important to reiterate that GazeBase has undergone prefiltering to an EXTRA level. 
This filtering approach, using a heuristic filter, effectively removes higher-frequency components. 
Consequently, applying a cutoff frequency of 75 Hz effectively distinguishes between ``signal'' and ``noise''. 
However, this approach should also be acknowledged as a limitation of the research. 
It indicates the need for an empirical study to evaluate biometric performance using unfiltered data.
In our research, we suggest some potential contributors to the noise component (e.g., data quality artifacts like glasses or eye color) which may explain our finding.  
This remains to be explored in future work.

\section{Conclusion}
We conducted a study where we compared the biometric performance of the ``signal'' and ``noise'' components of high-quality eye movement data originally recorded at 1000 Hz. 
Our research was validated using recordings from the publicly available GazeBase dataset \citep{Griffith2020}. To obtain biometric results, we used the state-of-the-art EKYT architecture \citep{lohr2022eye}.
Our findings indicated that the ``signal'' portion of the data performed well in both short-term and long-term data authentication, with relatively low EER, FRR @ FAR $10^{-4}$ and high d'. 
On the other hand, the ``noise'' part performs substantially better than chance when authenticating on short-term data but its performance was not as effective as the ``signal'' part.
We want to conclude by saying that the ``noise'' part carries individual-specific information that distinguishes between individuals. 
Thus, both the ``signal'' and ``noise'' parts of the recordings may have important biometric implications as well as privacy concerns.
Our findings also highlight the need for future studies using purely unfiltered data collected at 1000 Hz or lower sampling rate to assess the true potential of ``signal'' and ``noise'' separation.
A taxonomy of noise would be an important contribution to the eye-tracking field.  It would describe different parts of the noise, and how these parts form patterns that can be classified. 
This will be a significant undertaking.

\section*{Privacy and Ethics Statement}

This study promotes fairness and accessibility in eye-tracking across virtual, augmented, and mixed realities, adhering to stringent privacy and ethical standards. 
We prioritize respecting participant dignity, privacy, and rights and handle all data with utmost responsibility and confidentiality.

\bibliographystyle{unsrtnat}
\bibliography{ref}

\end{document}